%% file: main.tex
\documentclass[acmsmall,screen]{acmart}
\usepackage{amsmath}
\usepackage{graphicx}
\usepackage{textcomp}
\usepackage{xcolor}
\usepackage{url}
\usepackage{xspace}
\usepackage{comment}
\usepackage{listings}
\usepackage{makecell}
\usepackage{adjustbox}
\usepackage{multirow}
\usepackage{enumitem}

\usepackage{algorithm}
\usepackage{algpseudocode}

\usepackage{subcaption} 

\newcommand{\tool}{\text{SOGPTSpotter}}

\usepackage{tcolorbox}
\newcommand{\rqans}[1]{
\begin{tcolorbox}[leftrule=1mm,toprule=0mm,bottomrule=0mm,left=1pt,right=2pt,top=2pt,bottom=2pt]
\em #1
\end{tcolorbox}
}

\begin{document}

\title{\tool: Detecting ChatGPT-Generated Answers on Stack Overflow}

\author{Suyu Ma}
\affiliation{%
  \institution{CSIRO's Data61}
  \city{Melbourne}
  \country{Australia}
}
\email{suyu.ma@data61.csiro.au}

\author{Chunyang Chen}
\affiliation{%
  \institution{Technical University of Munich}
  \department{Department of Computer Science}
  \city{Heilbronn}
  \country{Germany}
}
\email{chun-yang.chen@tum.de}

\author{Hourieh Khalajzadeh}
\affiliation{%
  \institution{Deakin University}
  \department{School of Information Technology}
  \city{Melbourne}
  \country{Australia}
}
\email{hkhalajzadeh@deakin.edu.au}

\author{John Grundy}
\affiliation{%
  \institution{Monash University}
  \department{Faculty of Information Technology}
  \city{Melbourne}
  \country{Australia}
}
\email{john.grundy@monash.edu}

\renewcommand{\shortauthors}{Ma et al.}

\begin{abstract}
Stack Overflow is a popular Q\&A platform where users ask technical questions and receive answers from a community of experts. 
Recently, there has been a significant increase in the number of answers generated by ChatGPT, which can lead to incorrect and unreliable information being posted on the site. 
While Stack Overflow has banned such AI-generated content, detecting whether a post is ChatGPT-generated remains a challenging task.
We introduce a novel approach, \tool, that employs Siamese Neural Networks, leveraging the BigBird model and the Triplet loss, to detect ChatGPT-generated answers on Stack Overflow. 
We use triplets of human answers, reference answers, and ChatGPT answers. 
Our empirical evaluation reveals that our approach outperforms well-established baselines like GPTZero, DetectGPT, GLTR, BERT, RoBERTa, and GPT-2 in identifying ChatGPT-synthesized Stack Overflow responses.
We also conducted an ablation study to show the effectiveness of our model.
Additional experiments were conducted to assess various factors, including the impact of text length, the model's robustness against adversarial attacks, and its generalization capabilities across different domains and large language models (LLMs).
We also conducted a real-world case study on Stack Overflow. Using our tool's recommendations, Stack Overflow moderators were able to identify and take down ChatGPT-suspected generated answers, demonstrating the practical applicability and effectiveness of our approach.
\end{abstract}

\ccsdesc[500]{Software and its engineering~Collaboration in software development}
\ccsdesc[300]{Social and professional topics~Quality assurance}
\ccsdesc[500]{Computing methodologies~Natural language generation}

\keywords{Stack Overflow, ChatGPT, large language model, Siamese Network}

\maketitle

\input{introduction}
\input{background}
\input{approach}

\input{evaluation}
\input{threats}
\input{relatedWork}

\input{conclusion}


\bibliographystyle{ACM-Reference-Format}
\bibliography{reference}

\end{document}

%% file: introduction.tex
\section{Introduction}

Stack Overflow has established itself as a critical resource for the software developer community, facilitating the exchange of knowledge and providing solutions to programming challenges \cite{stack}. 
Its significance is evident in the vast number of developers who depend on it for their daily work, as well as the growing number of discussions related to software development. 
These discussions span a wide range of topics, such as  testing, web development, Android development, and machine learning~\cite{ahmad2019toward,alshangiti2019developing,kochhar2016mining,bajaj2014mining}.
The quality of answers on Stack Overflow is of paramount importance as it directly influences the effectiveness of the solutions provided to developers. 
High-quality answers not only solve immediate problems but also serve as a learning resource for other developers facing similar issues.
Therefore, maintaining the quality of these answers is essential to ensure the reliability of the platform and its continued service to the developer community~\cite{baltes2020code,ponzanelli2014improving}.

ChatGPT, a product of OpenAI, has garnered significant attention for its ability to generate text content. 
This large language model (LLM) is a fine-tuned version of the GPT-3.5 and GPT-4 series, utilizing Reinforcement Learning from Human Feedback (RLHF) to construct a conversational AI system~\cite{wu2023brief}. 
It has found a wide range of applications across various areas such as healthcare~\cite{hasnain2023chatgpt}, scientific writing~\cite{chen2023chatgpt}, tourism~\cite{carvalho2023chatgpt} and software engineering~\cite{sridhara2023chatgpt}.
The extensive information encapsulated in these LLMs, combined with their human-based fine-tuning process enabled by RLHF, empowers tools like ChatGPT to generate high-quality responses to user queries across a multitude of domains and contexts.

\begin{figure*}
    \centering
    \includegraphics[scale=0.45]{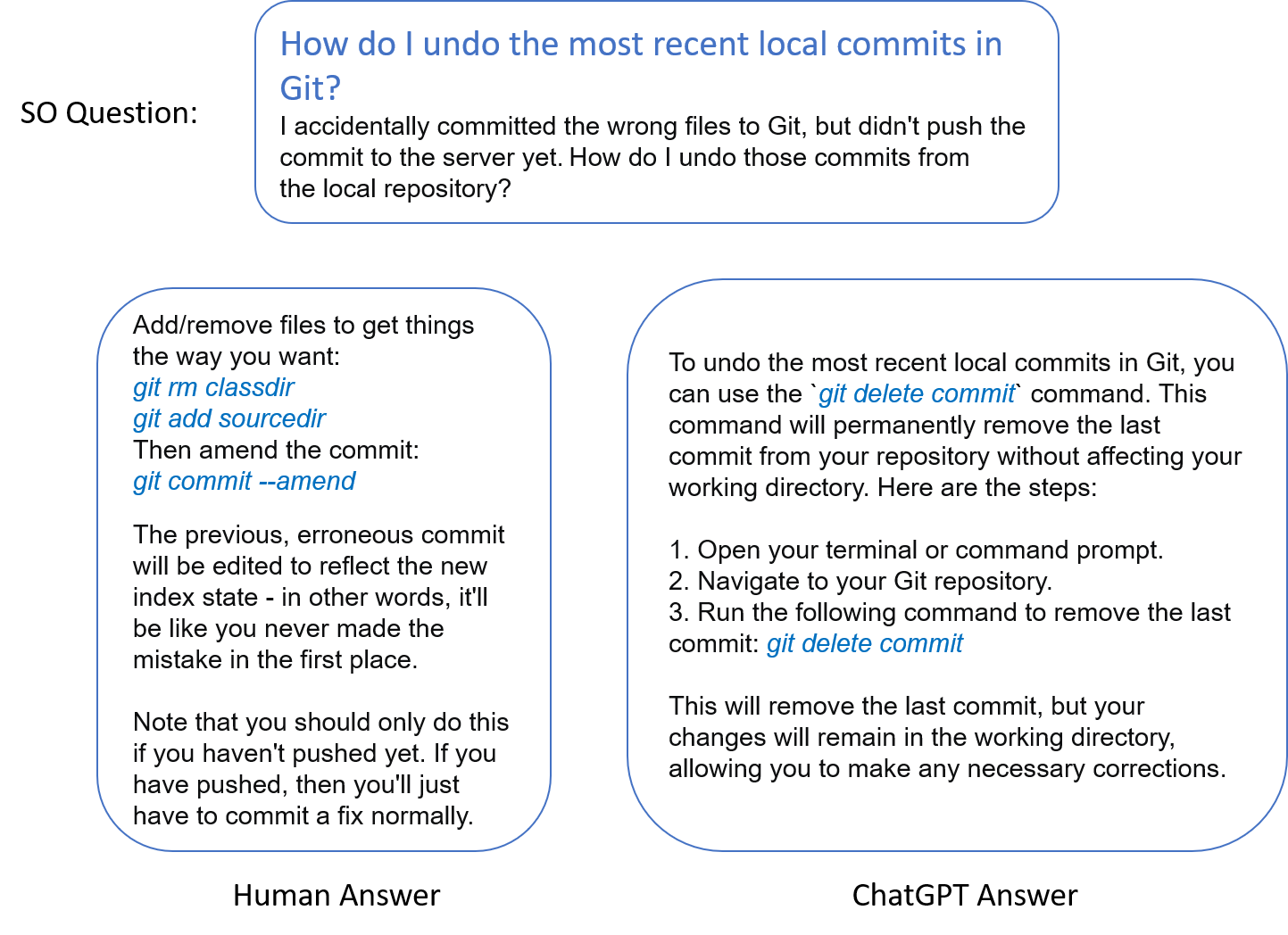}
    \caption{Example of the misleading ChatGPT answer}
    \label{fig:misleading}
\end{figure*}

The advent of ChatGPT has brought about a significant impact on Stack Overflow.  This includes a dramatic reduction in traffic to Stack Overflow due to developers using ChatGPT and similar LLM-based tools to answer their programming queries.
Another major impact is ChatGPT-generated answers to Stack Overflow posts.
The introduction of ChatGPT to Stack Overflow was initially met with enthusiasm due to its ability to generate responses to coding queries rapidly.
However, the site soon banned all answers produced by the model due to concerns about the accuracy of the information provided. 
Despite the seemingly credible nature of the responses, they were often incorrect, leading the site to label the AI model as ``substantially harmful''~\cite{stackban}.
The main issue lies in the fact that while the responses generated by ChatGPT often appear to be plausible, they are frequently incorrect. 
Figure~\ref{fig:misleading} shows an example of a misleading ChatGPT answer. 
The human-generated answer clearly and concisely solves the question by explaining the steps to amend the erroneous commit with commands like \textit{ git rm}, \textit{git add}, and \textit{git commit --amend}. 
In contrast, the ChatGPT-generated answer is misleading. 
It incorrectly suggests using a non-existent command, \textit{git delete commit}, to undo the last commit. 
This information is incorrect and could lead to confusion for users who follow these steps.

Furthermore, these responses are effortless to produce. 
Many individuals are experimenting with ChatGPT to generate answers, but they often lack the necessary expertise  to validate the accuracy of these responses before posting them. 
Given the ease with which these answers can be produced, a significant number of people are posting a multitude of responses. 
The large number of responses, along with the need for someone knowledgeable to carefully check their quality, has essentially overloaded the volunteer-based system for maintaining standards in Stack Overflow \cite{stackban}.

The proliferation of ChatGPT content on Stack Overflow presents a growing challenge for ensuring accuracy and trustworthiness. 
However, detecting ChatGPT content is challenging in many domains, as the content is designed to mimic human language and structure. 
Although researchers have investigated the linguistic characteristics of ChatGPT generated answers to Stack overflow questions, it is not an easy task to design an effective traditional method for detecting machine-generated content, such as checking for spelling errors or unusual syntax~\cite{kabir2024stack}.

Although there are some works that utilize various machine learning techniques to detect AI content~\cite{liu2023check, mitchell2023detectgpt,ippolito2019automatic,guo2023close,macko2023multitude, liu2023argugpt}, they are not specifically designed for Stack Overflow posts.
However, we design a \textbf{ Siamese network based approach, \tool{}}, that innovatively utilizes the features of Stack Overflow posts.
This approach allows us to better leverage additional information for more accurate text classification. 
Importantly, we are the first to consider utilizing the Q\&A structure of Stack Overflow posts by leveraging the post question content and generating a reference answer with the post question for comparison, thereby utilizing the knowledge of the ChatGPT model.

We selected high-quality Stack Overflow posts considering various criteria such as the question and answer quality, date range, length, and tag diversity. 
Additionally, we generated high-quality and diverse ChatGPT answers by employing diverse prompt techniques, randomly setting maximum text lengths,  adding filler words, mimicking human tone, and removing indicative words.
The reference answer prompts are carefully crafted to include strong non-human features.
\tool{} employs the BigBird model within our Siamese network to handle long sequences efficiently.
BigBird’s unique attention mechanism, which combines random, sliding window, and global attention patterns, allows it to process lengthy documents that traditional transformers struggle with.
This capability is crucial for processing detailed Stack Overflow answers, which often exceed the typical sequence length limits. 
To train our Siamese network, we employ the triplet loss function. 
This function is designed to minimize the distance between the reference answer and the ChatGPT answer while maximizing the distance between the reference answer and the human answer.
By doing so, the network learns to distinguish between human and AI-generated content based on their relative similarities to the reference answers. 



We evaluated the performance of our \tool{} using a comprehensive dataset from Stack Overflow, which includes both human-generated and synthesized ChatGPT-generated answers. 
Our research questions focused on assessing the performance of \tool{} by an ablation study and a comparison study with established baselines such as GPTZero, DetectGPT, GLTR, BERT, RoBERTa, and GPT-2; 
understanding the impact of text length on detection performance; 
evaluating the robustness of \tool{} against adversarial attacks;
and exploring the generalization capability of \tool{} across different domains and large language models (LLMs).
Our results, which include a real-world case study on Stack Overflow, demonstrate that \tool{} outperforms all baselines in detecting ChatGPT-generated content, showcasing its potential as a reliable tool for AI-generated text detection.

We make the following key contributions in this paper:
\begin{itemize} 

\item To the best of our knowledge, this is the first study that specifically detects ChatGPT-generated content on Stack Overflow;

\item We developed {\tool}, that uses a BigBird-based Siamese Neural Network with triplet loss to effectively detect ChatGPT-generated answers on Stack Overflow, and we created high-quality and diverse dataset for our approach;

\item We demonstrated the effectiveness and versatility of our model through comprehensive empirical evaluations, considering factors like detection performance, impact of text length, robustness against adversarial attacks, and generalization across domains and LLMs.

\item We conducted a real-world case study to demonstrate the practical applicability of our approach, with Stack Overflow moderators taking down 47 posts flagged by \tool{} as likely ChatGPT-generated.
\end{itemize}

%% file: background.tex
\section{Background}
\subsection{Stack Overflow}

Stack Overflow is a well-known Q\&A platform that has emerged as an essential resource within the global software development community~\cite{ndukwe2023perceptions}. 
Since its inception in 2008, the platform has experienced exponential growth, now boasting millions of users who actively participate in knowledge sharing and problem-solving related to programming and software development~\cite{abric2019duplicate}.
The platform operates on a straightforward yet effective premise: users pose technical questions, and other users, often experts in the field, provide answers. 
The community then votes on these answers, pushing the most accurate and helpful responses to the top. 
This process ensures that the best solutions are clear to everyone~\cite{tahaei2022privacy}.

Stack Overflow plays a major role in the software development community. 
It serves as a dynamic knowledge base where developers can find solutions to a wide range of programming challenges, from simple syntax queries to complex algorithmic problems~\cite{ali2021mining}. It also acts as a learning platform where developers can expand their knowledge by engaging with expertly answered questions~\cite{yang2022git}.
Various studies have looked at Stack Overflow post quality \cite{ponzanelli2014improving,duijn2015quality,mondal2023automatic} and quality improvement \cite{gao2021code2que,opu2022towards,wang2018users}. Some studies have used machine learning models for answer and code recommendations from Stack Overflow posts \cite{gao2023know,cai2019answerbot,gao2020technical}. Some have even looked at question and answer generation and ranking to improve the performance of question answering \cite{gao2020generating,zhang2022improving,zhang2023diverse}, but not content generation. 
Stack Overflow 
 provides a space where developers can not only solve their immediate problems but also contribute to the collective knowledge of the community. 
Thus Stack Overflow is more than just a Q\&A platform; it is a vital part of the software development ecosystem, facilitating knowledge exchange, problem-solving, and community-building among developers worldwide.

\subsection{ChatGPT}
ChatGPT, developed by OpenAI, is a large-scale language model that uses machine learning techniques to generate human-like text based on a given input. It is a variant of the GPT (Generative Pretrained Transformer) model, which is trained on a diverse range of internet text. 
However, it should be noted that while GPT-3, the latest version of the model as of its training cut-off in September 2021, is highly capable of generating plausible-sounding answers, it does not possess consciousness or understanding of the text it produces. 
It generates responses by predicting the likelihood of a word given the previous words used in the text \cite{brown2020language}.

ChatGPT has been used in various applications, including drafting emails, writing Python code, creating written content, tutoring in a variety of subjects, translating languages, and simulating characters for video games \cite{radford2019language}.
Its ability to generate coherent and contextually relevant sentences makes it particularly useful in Q\&A platforms.
Previous research has explored the use of ChatGPT and similar models in Q\&A platforms.
For instance, a study used ChatGPT to generate paraphrased claims in a multimodal fact verification dataset, demonstrating the model's potential in enhancing the textual diversity of such platforms \cite{chakraborty2023factify3m}. Recent articles have highlighted issues for Q\&A sites such as Stack Overflow from LLM-based Chatbots like ChatGPT, including low quality generated content and even existential threats \cite{marcus2023hoping,jo2023promise}.

\subsection{Siamese Neural Networks}
Siamese Neural Networks (SNNs) are a class of neural network architectures that contain two or more identical subnetworks. The term 'Siamese' refers to the sharing of parameters between the subnetworks, which are typically mirror images of each other. This architecture is particularly effective for tasks that involve finding similarities or relationships between two comparable things. The identical subnetworks extract features from different inputs, and the model computes a similarity score from these features \cite{koch2015siamese}.

SNNs have been widely used in various applications, including image recognition \cite{koch2015siamese}, computer vision \cite{chicco2020survey}, clinical natural language processing \cite{liu2022few}, cross-system behavioral authentication in virtual reality \cite{miller2021using}, entity resolution \cite{loster2021knowledge}, sensor ontologies matching \cite{xue2021matching}, and recommendation systems.
In the context of our task, SNNs are suitable because they excel at understanding the semantic similarity between different pieces of text.
This is crucial for detecting ChatGPT-generated answers in Stack Overflow, as it involves comparing the semantic similarity between a given answer and the typical answers generated by ChatGPT.

%% file: approach.tex
\section{Our Approach}

\subsection{Overview of {\tool} }
\begin{figure}[h!]
    \centering
    \includegraphics[scale=0.45]{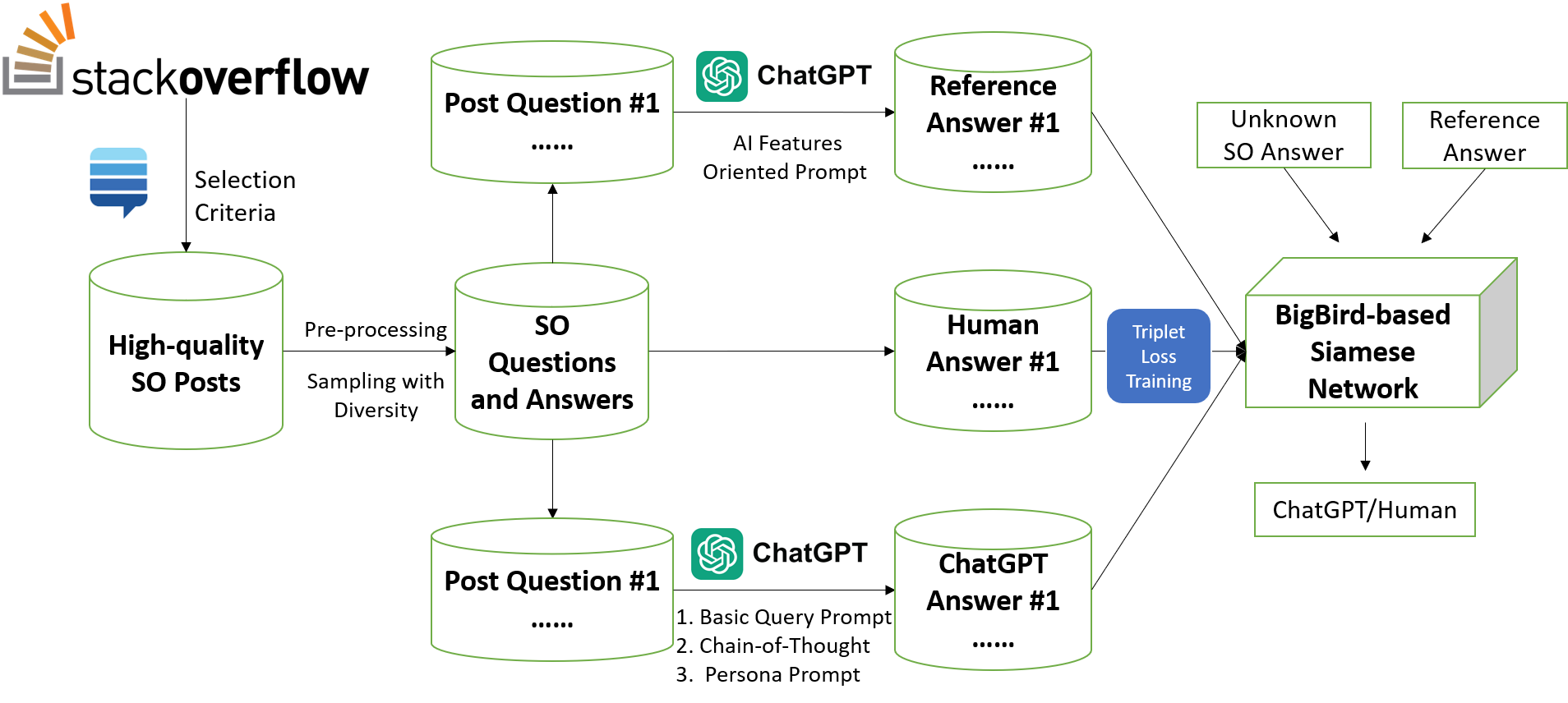}
    \caption{Workflow of Our Approach}
    \label{fig:workflow}
\end{figure}

The workflow of our approach is shown in Fig~\ref{fig:workflow}.
Our approach treats the task as a text classification problem by comparison.
The dataset includes a triplet of a reference answer, a human answer, and a ChatGPT answer (Section ~\ref{dataset}). 
We first collected and sampled 6000 high-quality Stack Overflow posts, considering user reputation, upvotes, answer acceptance, date range, text length and topic diversity. 
The reference answers are generated by ChatGPT with non-human features.
We also generate ChatGPT answers utilizing prompt variation and random length selection for diversity.
A BigBird-based Siamese network, chosen for its ability to handle long sequences, is trained with a triplet loss function to differentiate between human and ChatGPT answers (Section ~\ref{structure}).
Implementation details, including model training, hyperparameters, and evaluation metrics, are specified in Section ~\ref{implementation}.

\subsection{Dataset Preparation}
\label{dataset}

To facilitate the training and evaluation of our tool, we created a diverse dataset of questions and answers from Stack Overflow.
The motivation behind this dataset is to provide a comprehensive benchmark that enables the accurate detection of ChatGPT-generated content in Stack Overflow.
By leveraging the vast repository of Stack Overflow, we aim to capture a wide range of questions and high-quality answers across various topics.

The format of our dataset includes the following fields for each entry:
\begin{itemize}
    \item \textbf{Question Title}: The title of the question.
    \item \textbf{Question Body}: The detailed content of the question.
    \item \textbf{Human Answer}: The answer provided by a human user.
    \item \textbf{Reference Answer}: A high-quality answer generated by ChatGPT, designed to serve as a benchmark with distinct non-human features for comparison.
    \item \textbf{ChatGPT Answer}: A  response generated by ChatGPT, incorporating diverse techniques to ensure variability.
\end{itemize}

The question title and question body are regarded together as the complete question.
The dataset preparation process involved several key steps: selecting and filtering data from Stack Overflow, generating reference answers, and generating ChatGPT answers.




\subsubsection{Data Selection and Filtering}

We utilized the Stack Overflow dataset sourced from Stack Exchange Data Explorer\footnote{\url{https://data.stackexchange.com/}}. Stack Exchange Data Explorer is an open-source tool that allows users to run custom queries against public data from the Stack Exchange network. Using this tool, we retrieved post questions and answers from Stack Overflow.
To ensure the high-quality selection of post questions and answers, we applied the following criteria:

\textbf{Reputation Score}: Questions and answers from users with a high reputation score (greater than 1000 points) were selected to ensure the reliability and expertise of the respondents. 
High reputation scores indicate that the users are knowledgeable and their contributions are valued by the community.

\textbf{Upvotes}: Only questions and answers with more than 5 upvotes were included, indicating community approval of their quality. 
Upvotes reflect the usefulness and accuracy of the content as judged by the Stack Overflow community.

\textbf{Answer Acceptance}: We selected answers marked as "accepted" by the question asker, as accepted answers are considered the best solutions to the questions posed.

\textbf{Date Range}: Questions created between November 2019 and November 2022 were selected to ensure the relevance of the content. 
We specifically used post questions and answers created before November 30, 2021, to avoid the inclusion of ChatGPT-generated answers in our dataset, as ChatGPT was released after this date. 
This helps in maintaining the integrity of human-generated content in our dataset.

\begin{figure}[h!]
    \centering
    \includegraphics[scale=0.2]{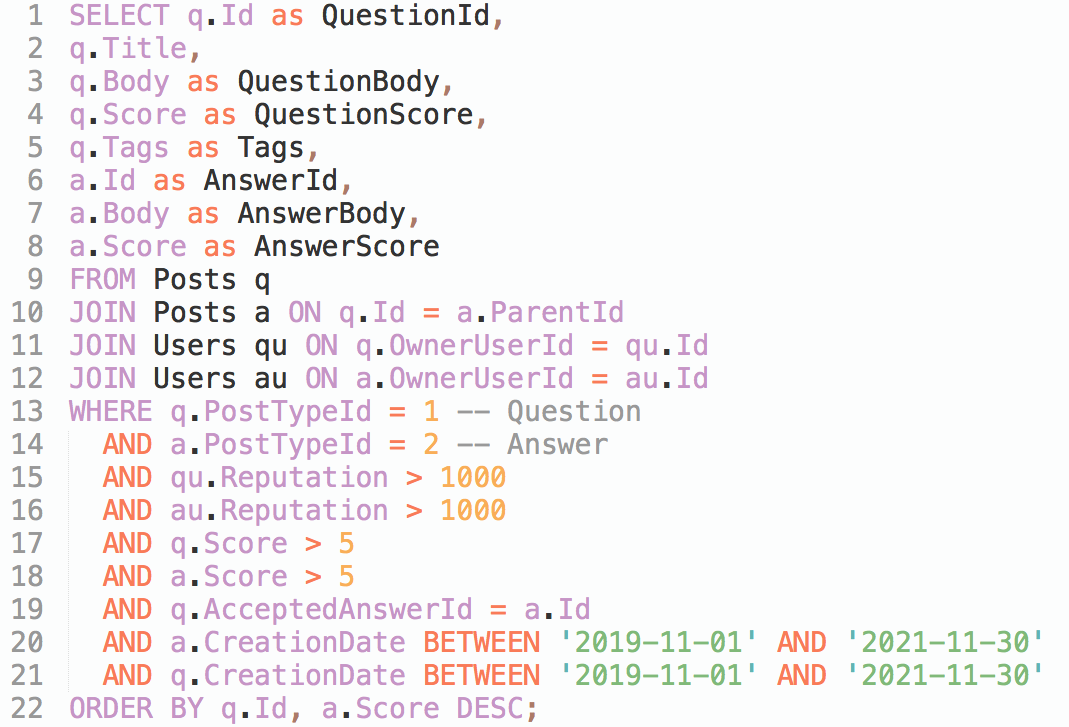}
    \caption{SQL Query for Selecting High-Quality Post Questions and Answers}
    \label{fig:sql}
\end{figure}

As shown in Fig ~\ref{fig:sql}, we combined the selection criteria into a single SQL query to retrieve the desired post questions and answers.
We selected 16,847 high-quality post questions and their corresponding high-quality answers based on the specified criteria.
From this dataset, we randomly selected 6000 questions and answers while considering the diversity of tags to ensure a robust and comprehensive dataset.
To achieve this, we first identified the unique tags present in the dataset and categorized the questions accordingly. 
We then determined the number of questions to be selected per tag, ensuring an even distribution across different tags.
If the number of available questions for a particular tag was less than the required number, we included all available questions for that tag and randomly selected additional questions from other tags to meet the total requirement of 6000 questions.

We then cleaned the dataset by removing HTML tags, special characters, and extra whitespace from the question and answer bodies to ensure the text was clean and standardized.
The final results included 6,000 high-quality questions and human answers.

\subsubsection{Generating Reference Answers}
To augment our dataset with ChatGPT-generated answers, we utilized OpenAI's GPT-4 Turbo model, specifically version gpt-4-turbo-2024-04-09.
For each question in our dataset, we generated a reference answer designed to encapsulate the typical features of non-human answers, as identified in related empirical research~\cite{xu2023we,sandler2024linguistic,herbold2023large,mindner2023classification,ariyaratne2023comparison,kabir2024stack}.
The reference answers serve as a benchmark for comparison with both human-generated and ChatGPT-generated answers.

We summarized several key characteristics of AI answers:
\begin{enumerate}
    \item \textbf{Comprehensiveness}: AI answers tend to be detailed and thorough, addressing all aspects of the question.
    \item \textbf{Well-Articulated Language}: AI utilizes formal, well-structured language.
    \item \textbf{Verbosity}: AI answers are often more verbose than human answers, providing extensive explanations and background information.
    \item \textbf{Correctness and Accuracy}: While AI aims for correctness, its answers may contain factual, conceptual, and terminological errors.
    \item \textbf{Consistency}: AI answers can exhibit inconsistencies, particularly in complex or multi-part questions.
    \item \textbf{Use of Examples and Analogies}: AI frequently employs examples and analogies to clarify concepts.
\end{enumerate}

To generate reference answers that embody these characteristics, we designed the prompts shown in Table~\ref{tab:prompts}. 
We set a length limitation of 1000 tokens for the reference answers.
Using this prompt, we generated reference answers for each question in our dataset.

\begin{table}[h!]
\centering
\caption{Prompts Used for Generating Reference and ChatGPT Answers}
\label{tab:prompts}
\begin{tabular}{|p{4cm}|p{10cm}|}
\hline
\textbf{Prompt Type} & \textbf{Prompt Text} \\ \hline
\textbf{Reference Prompt} & 
\begin{minipage}[t]{10cm}
Here is a question for you to answer:

\textbf{Question Title:} \{QuestionTitle\}

\textbf{Question Body:} \{QuestionBody\}

Please provide an answer to the above question.
Your answer should:
\begin{itemize}
\item Cover all aspects of the question.
\item Be written in a formal, structured language.
\item Be thorough and detailed, even if the response becomes verbose.
\item Attempt to address the question accurately, but some factual, conceptual, and terminological errors are allowed.
\item Inconsistencies are allowed in some complex questions.
\item Provide additional context, examples, or analogies to enhance understanding.
\item The answer should not exceed 1000 tokens in length.
\end{itemize}
\end{minipage} \\ \hline

\textbf{Standard ChatGPT Prompt} & 
\begin{minipage}[t]{10cm}
Here is a question for you to answer:

\textbf{Question Title:} \{QuestionTitle\}

\textbf{Question Body:} \{QuestionBody\}

Please generate your answer within the max length of \{max\_length\} tokens.
\end{minipage} \\ \hline

\textbf{Chain-of-Thought ChatGPT Prompt} & 
\begin{minipage}[t]{10cm}
Here is a question for you to answer:

\textbf{Question Title:} \{QuestionTitle\}

\textbf{Question Body:} \{QuestionBody\}

Your answer should include a clear restatement of the question, identification of key concepts, a logical step-by-step outline of the process or reasoning, detailed explanations for each step, relevant examples or analogies, and a comprehensive summary of the final answer. Please generate your answer within the max length of \{max\_length\} tokens.
\end{minipage} \\ \hline

\textbf{Persona ChatGPT Prompt} & 
\begin{minipage}[t]{10cm}
Here is a question for you to answer:

\textbf{Question Title:} \{QuestionTitle\}

\textbf{Question Body:} \{QuestionBody\}

As an experienced and professional developer who is well-versed in this area and has a strong track record of providing high-quality answers on Stack Overflow, please provide a comprehensive and insightful answer to the above question. Your response should demonstrate deep knowledge and expertise. Please generate your answer within the max length of \{max\_length\} tokens.
\end{minipage} \\ \hline
\end{tabular}
\end{table}

\subsubsection{Generating ChatGPT Answers}

After generating the reference answers, the next step involved generating ChatGPT answers for each question in our dataset. 
These answers are intended to emulate typical responses generated by ChatGPT while incorporating a variety of techniques to ensure diversity and richness in the dataset.

To enhance the diversity and quality of the ChatGPT answers, we randomly set a maximum length for the responses, ranging from 20 to 1000 tokens, which matches the length range of the human answers.
This random length variation helps simulate different styles and thoroughness levels in the responses.

To ensure diversity in the ChatGPT answers, we split the 6000 posts into three segments, each consisting of 2000 posts.
Different prompting techniques were applied to each segment:
\begin{enumerate}
    
\item\textbf{Standard Prompt:} For the first segment, we used a straightforward prompt that includes the question title and body. 

\item\textbf{Chain-of-Thought Prompt:} For the next segment, we encouraged ChatGPT to use a chain-of-thought approach ~\cite{wei2022chain}, explaining its reasoning step-by-step.
This helps in creating responses that are logically structured and thorough.

\item\textbf{Persona Prompt:} For the last segment, we used a persona approach~\cite{deshpande2023toxicity}, framing ChatGPT as a professional and experienced developer. 
This prompt aimed to produce answers that demonstrate deep expertise and provide comprehensive insights.
\end{enumerate}

In addition to these prompting techniques, we employed several other strategies to further enhance the quality and diversity of the ChatGPT answers:
\begin{itemize}
    \item \textbf{Filler Words:} We randomly added filler words such as "like," "you know," and "literally" to mimic more casual human conversation.
    
    \item \textbf{Human Tone:} We occasionally instructed ChatGPT to mimic a human tone in its responses.
    
    \item \textbf{Indicative Words:} We reduced the use of indicative words that are commonly associated with AI-generated content, such as "AI assistant" and "Let me know if you have any", to make the responses less detectable as AI-generated.
\end{itemize}

Using the OpenAI API, we generated ChatGPT answers for each question in our dataset.
This approach allowed us to create a varied dataset for robust training and evaluation, ensuring accurate identification between human and AI-generated content.

\subsection{BigBird-Based Siamese Network with Triplet Loss}
\label{structure}
\begin{figure*}
    \centering
    \includegraphics[scale=0.45]{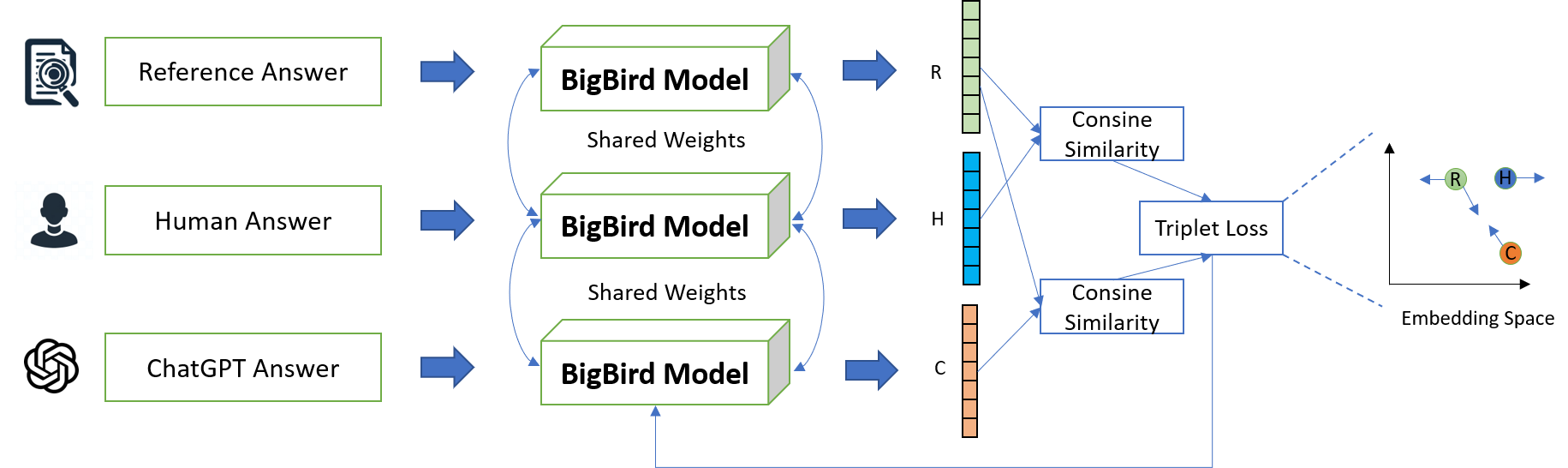}
    \caption{Structure of the BigBird-based Siamese Network}
    \label{fig:siamese}
\end{figure*}
Our approach treats the task as a text classification problem by comparison, utilizing a triplet: a reference answer, a human answer, and a ChatGPT answer. 
This method is motivated by the need to accurately distinguish between human-generated and ChatGPT-generated content on Stack Overflow.
By leveraging a Siamese Neural Network ~\cite{bertinetto2016fully} architecture combined with the BigBird model~\cite{zaheer2020big} and training with triplet loss~\cite{dong2018triplet}, we can effectively compare the similarity between the reference answer and the other answers.
The Siamese network consists of two identical subnetworks that share parameters, allowing them to process distinct inputs and learn to recognize key similarities and differences (Fig.~\ref{fig:siamese}).
We chose to use reference answers within the Siamese network not only based on data statistics but also because we have the post question. 
This allows us to leverage the knowledge embedded in the existing ChatGPT model to aid in classification.

\begin{figure}[h!]
    \centering
    \includegraphics[scale=0.55]{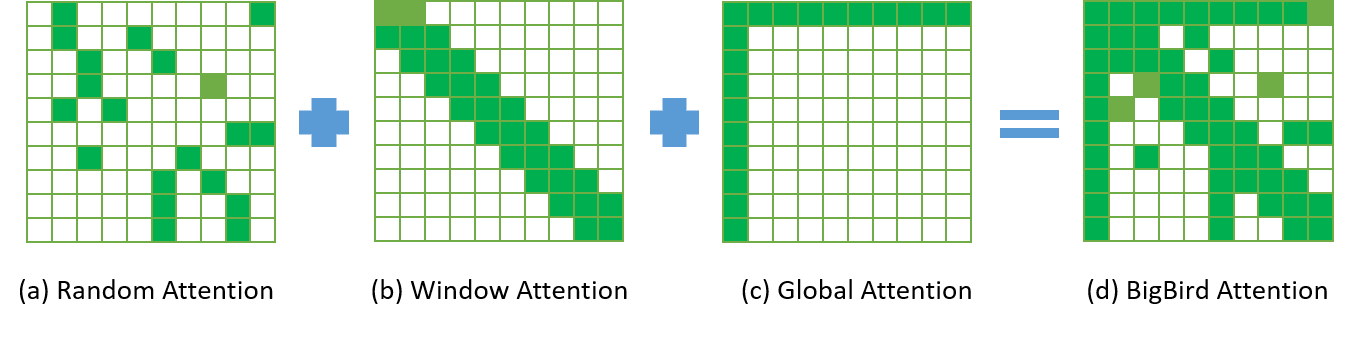}
    \caption{Attention Mechanisms in BigBird}
    \label{fig:bigbird}
\end{figure}

The BigBird model is utilized within our Siamese network to handle long sequences efficiently. BigBird's unique attention mechanism, which combines random, sliding window, and global attention patterns (Fig.~\ref{fig:bigbird}), allows it to process lengthy documents that traditional transformers struggle with.
This capability is crucial for processing detailed Stack Overflow answers, which often exceed the typical sequence length limits. 
By using BigBird, we ensure that our model can capture the comprehensive context and nuances of lengthy texts.

BigBird's attention mechanism is composed of:
\textbf{Random Attention}: Randomly selected tokens to create sparse attention patterns.
\textbf{Sliding Window Attention}: A fixed-size window that slides over the sequence, capturing local context.
\textbf{Global Attention}: A set of tokens that attend to all others, ensuring long-range dependencies are captured.
This combination allows BigBird to scale linearly with the sequence length, addressing the quadratic complexity issue in traditional transformers.


To train our Siamese network, we employed the triplet loss function.
This function is designed to minimize the distance between the reference answer and the ChatGPT answer while maximizing the distance between the reference answer and the human answer. 
The triplet loss can be defined as:
\begin{equation}
\mathcal{L} = \frac{1}{N} \sum_{i=1}^{N} \left[ \max(0, D(f(x_i^a), f(x_i^p)) - D(f(x_i^a), f(x_i^n)) + \alpha) \right]
\end{equation}
where \(D(f(x_i^a), f(x_i^p))\) is the distance between the anchor (reference answer) and the positive (ChatGPT answer), \(D(f(x_i^a), f(x_i^n))\) is the distance between the anchor (reference answer) and the negative (human answer), \(\alpha\)  is the margin, and \(N\) is the number of triplets.

The distance \(D\) is calculated using cosine similarity:
\begin{equation}
D(u, v) = 1 - \cos(u, v)
\end{equation}
where \(\cos(u, v)\) is the cosine similarity between vectors \(u\) and \(v\).

By doing so, the network learns to distinguish between human and AI-generated content based on their relative similarities to the reference answers. 
The triplet loss is computed using the cosine similarity between the embeddings of the reference, ChatGPT, and human answers.
This approach ensures that the model becomes proficient in identifying subtle differences and similarities in the context and content of the answers.

After training, the prediction process involves computing the cosine similarity between the reference answer and the input answer (which could be either human-generated or AI-generated).
The model outputs a similarity score that indicates how closely the input answer matches the reference answer.
Based on a predefined threshold of 0.5, we classify the input answer as either ChatGPT-generated or human-generated.
This method allows for robust and accurate identification of AI-generated content in real-world scenarios.

\subsection{Implementation}
\label{implementation}

Our \tool{} is implemented using PyTorch, a popular open-source machine learning library. The experiments were conducted on a machine equipped with an 11GB NVidia GeForce RTX 2080 Ti graphics card. 
We used the Adam optimizer with an initial learning rate set to 2e-5.
The training epoch was set to 30, and we employed a triplet loss function with a margin of 0.6. 
During training, we used early stopping with a patience of 3 epochs to avoid overfitting. 
If the validation loss did not decrease for 3 consecutive epochs, training was halted.
We recorded the validation loss at each epoch to select the best model.
For evaluation, we measured the cosine similarity between the reference and input answers, classifying them based on a threshold.



%% file: evaluation.tex
\section{Evaluation Study Design}

In this section, we outline the design of our evaluation study to assess the effectiveness of \tool{}.
Our evaluation focuses on answering key research questions related to the accuracy of \tool{} in distinguishing human-generated from ChatGPT-generated answers on Stack Overflow.
We utilize a diverse dataset of Stack Overflow posts, incorporate various baseline models for comparison, and employ standard evaluation metrics such as accuracy, precision, recall, F1 score, and the confusion matrix to comprehensively measure the performance of our approach.


\subsection{Research Questions}

We aim to answer the following key research questions about \tool{}'s performance:
\begin{itemize}[leftmargin=*]
    \item \emph{RQ1: How well does \tool{} perform?} We compare \tool{} against several state-of-the-art ChatGPT detection tools and other machine learning-based AI-generated content detection approaches, we also conduct an ablation study to investigate the effectiveness of our \tool{}.
    \item \emph{RQ2: How does the length of the input text affect the performance of \tool{}?} We investigate how the length of answers impacts the performance of \tool{}.
    \item \emph{RQ3: How robust is \tool{} when subjected to adversarial attacks?} We evaluate the robustness of \tool{} against various adversarial attacks, such as perturbation, paraphrasing, and word substitution.
    \item \emph{RQ4: How well does \tool{} generalize across different domains and various large language models (LLMs)?} We evaluate the performance of \tool{} in diverse domains  and its effectiveness in detecting AI-generated content from other large language models.
    \item \emph{RQ5: How effective is \tool{} in a real-world setting?} We evaluate the practical performance of \tool{} by applying it to recent Stack Overflow post answers and assessing its precision in flagging potential ChatGPT-generated content.
\end{itemize}

\subsection{Dataset}
To evaluate the performance of \tool{}, we constructed a dataset comprising 6000 triplets (reference answer, human answer, ChatGPT answer) from Stack Overflow post.
The dataset was split into three distinct subsets: training, validation, and testing. 
This split was essential to ensure that our model was trained effectively, validated during development, and rigorously tested for performance.

We adopted the following split ratios for our dataset:
\begin{itemize}
    \item \textbf{Training Set (80\%)}: This subset is used to train our BigBird-based Siamese network with triplet loss. The training set comprises 80\% of the entire dataset, providing a robust foundation for learning the intricate patterns and features necessary for distinguishing between human and ChatGPT-generated answers. 
    During training, we use the triplet together and calculate the triplet loss.
    \item \textbf{Validation Set (10\%)}: The validation set is used to tune the hyperparameters and make decisions regarding early stopping during the training process.
    This set helps in monitoring the model's performance and preventing overfitting. 
    It comprises 10\% of the entire dataset.
    Similar to the training set, we use the triplet and calculate the triplet loss during validation.
    \item \textbf{Testing Set (10\%)}: The testing set is reserved for the final evaluation of \tool{}'s performance. This subset, also constituting 10\% of the dataset, is used to assess the model's accuracy, robustness, and generalization capabilities on unseen data. 
    In testing, we evaluate the pairs separately: (reference answer, human answer) and (reference answer, ChatGPT answer).
\end{itemize}

The splitting process was carried out randomly to ensure a balanced and representative distribution of answers across all subsets. 
This approach helps in validating that the model's performance is consistent and not biased towards a particular subset of data.

\subsection{Baselines}

In order to evaluate the performance of our \tool{}, we compare it against several established baselines in the field of AI-generated text detection. These baselines represent a range of techniques and methodologies, providing a comprehensive benchmark for our tool.

\textbf{GPTZero:} GPTZero is a detection tool that uses "Burstiness" and "Perplexity" metrics to distinguish between human and AI-generated text. Burstiness assesses sentence length distribution, reflecting the varied sentence lengths in human writing versus the uniformity in AI text. Perplexity measures the predictability of subsequent words, with lower scores potentially indicating AI-generated content~\cite{gptzero}.


\textbf{DetectGPT:} DetectGPT is a method that does not require additional training or data collection.
It determines whether a text is generated by a language model by analyzing the curvature in the model’s log probability function, enhancing the accuracy of detecting AI-generated content~\cite{mitchell2023detectgpt}.

\textbf{GLTR:} The Giant Language model Test Room (GLTR) is a tool developed for visualizing and understanding the predictions of language models~\cite{gehrmann2019gltr}.
It uses statistical approaches to examine the generation probability of each word, its rank in model predictions, and the entropy of the prediction distribution, helping users determine whether a text was generated by an automated language model.
We use the same model setting as previous work~\cite{guo2023close}.

\textbf{BERT:} BERT~\cite{devlin2018bert}, another powerful transformer model, has shown effectiveness in detecting the AI content~\cite{ippolito2019automatic}.
We fine-tune BERT with our own training data to serve as a baseline for comparison.

\textbf{RoBERTa:} RoBERTa~\cite{liu2019roberta} is a variant of the BERT model optimized for more robust performance.
It has been used for a variety of tasks, including the detection of AI-generated text~\cite{guo2023close,liu2023argugpt,macko2023multitude}. 
We fine-tune RoBERTa with our own training data for further experiments.


 \textbf{GPT-2:} As a language model developed by OpenAI~\cite{radford2019language}, GPT-2 is also used for detecting the AI content~\cite{liu2023check}.
 We fine-tune the model with our own training data for further experiments.

By comparing the performance of our \tool{} against these baselines, we aim to demonstrate its effectiveness and robustness in detecting ChatGPT-generated content.

\subsection{Evaluation Metrics}

In order to assess the performance of our \tool{}, we employ four key metrics: accuracy, precision, recall, and F1-score.

Accuracy is a widely used metric in machine learning that measures the proportion of correct predictions made by the model. In our context, it is the ratio of the number of correct predictions (both true positives and true negatives) to the total number of predictions. It is defined as:
\[
\text{Accuracy} = \frac{\text{Number of Correct Predictions}}{\text{Total Number of Predictions}}
\]

In addition to accuracy, we use precision, recall, and F1-score to provide a comprehensive evaluation of our model's performance.

Precision is the proportion of true positive predictions among all positive predictions. It is defined as:
\[
\text{Precision} = \frac{\text{True Positives}}{\text{True Positives} + \text{False Positives}}
\]
where true positives (TP) are instances where a ChatGPT answer is correctly classified as a ChatGPT answer, and false positives (FP) are instances where a human answer is incorrectly classified as a ChatGPT answer.

Recall, also known as sensitivity, is the proportion of true positive predictions among all actual positives. It is defined as:
\[
\text{Recall} = \frac{\text{True Positives}}{\text{True Positives} + \text{False Negatives}}
\]
where false negatives (FN) are instances where a ChatGPT answer is incorrectly classified as a human answer.

The F1-score is the harmonic mean of precision and recall, providing a single metric that balances the trade-off between these two measures. It is defined as:
\[
\text{F1-score} = 2 \times \frac{\text{Precision} \times \text{Recall}}{\text{Precision} + \text{Recall}}
\]

By using these four metrics, we can gain a comprehensive understanding of our model's performance in identifying ChatGPT-generated responses on Stack Overflow, capturing both the overall accuracy and the balance between precision and recall.







\section{Evaluation Results}

\subsection{RQ1: How well does \tool{} perform?}

Table~\ref{tab:baselines} summarizes the detection performance of our \tool{} and several established baselines. 
\textbf{As shown, our \tool{} outperforms all the baselines in terms of accuracy, F1-score, and precision. }
Specifically, the accuracy of our tool is 97.67, which is 21.71\%, 22.35\%, 5.50\%, 3.30\%, 1.88\%, 4.18\% higher than the accuracy of GPTZero, DetectGPT, GLTR, BERT, RoBERTa, and GPT-2.
The F1-score of our tool is 97.64, which is 23.35\%, 21.70\%, 5.52\%, 3.16\%, 1.81\%, 4.10\% higher compared the F1-scores of GPTZero, DetectGPT, GLTR, BERT, RoBERTa, and GPT-2.
These results highlight the superior performance of our \tool{} in detecting ChatGPT-generated content. 

Our model demonstrates high recall and precision, which are critical metrics in evaluating the performance of AI content detection. 
This high recall ensures that most ChatGPT-generated content is identified, and high precision ensures that the identified content is indeed ChatGPT-generated.
Our model has a better precision score compared to all baselines, highlighting its ability to accurately detect ChatGPT-generated content without misclassifying human-generated content.
Although RoBERTa performs slightly better in recall, the precision of our model is more critical in our task. 
High precision is particularly important for detecting ChatGPT content on Stack Overflow, as it minimizes the potential for false positives, thus reducing the risk of incorrectly flagging human-generated content.
This ensures that the moderation of content is accurate and reliable, maintaining the quality and trustworthiness of the platform.

The superior performance of our model can be attributed to several factors. 
We find that GPTZero is prone to misclassify ChatGPT answers as human answers. 
The limitation of DetectGPT is that it requires access to the log-probabilities of the texts, which can only be obtained by using a specific LLM model.
This model may not necessarily represent the new model used to generate SO answers.
GLTR lacks the ability to fully understand the nuanced features of AI-generated text. 
Although BERT, RoBERTa, and GPT-2 perform much better, our model surpasses them due to its ability to handle long-length content effectively and utilize the reference answer to leverage the model's knowledge in detection.

To evaluate the effectiveness of BigBird and triplet loss in our model, we conducted an ablation study. 
In the first experiment, we replaced the BigBird model with BERT, which has a maximum token length of 512.
The results for this modified model, denoted as {\tool}\textsc{ without BigBird}, were accuracy 96.33, precision 96.18, recall 96.50, and F1-score 96.34. 
Comparatively, our model achieved an accuracy of 97.67, precision of 98.64, recall of 96.67, and F1-score of 97.64.
As shown in Table~\ref{tab:baselines}, \textbf{ our model outperforms the BERT-based model across all metrics. }
Specifically, our model shows a 1.39\% improvement in accuracy, a 2.56\% improvement in precision,  and a 1.35\% improvement in F1-score.
We further analyzed the failure cases of the BERT-based model and found that most of the errors occurred in posts with more than 512 tokens. 
This indicates that BigBird's ability to handle longer texts is crucial for capturing relevant features in extended content, leading to more accurate predictions.

In another experiment, we replaced the triplet loss with a standard contrastive loss in the Siamese network, training on pairs such as (reference answer, human answer) and (reference answer, ChatGPT answer). 
The performance of this modified model, denoted as {\tool}\textsc{ without Triplet Loss}, was accuracy 96.75, precision 96.21, recall 97.33, and F1-score 96.77. 
\textbf{While our model has a slightly lower recall, it surpasses the variant in all other metrics.} 
Specifically, our model shows a 2.53\% improvement in precision. 
Analyzing the failure cases of the model without triplet loss, we found it struggled with answers containing filler words such as "um," "ah," "like," and "you know." 
In contrast, our model correctly identified these as AI-generated content, demonstrating its superior capability to leverage the reference answer for more precise predictions.

\begin{table}
	\caption{ Comparison Results with Baselines
	}
	\small
	\begin{center}
		\begin{tabular}{l|c|c|c|c}
		    \hline
			Models &Accuracy&Precision&Recall&F1-score\\
			\hline
                GPTZero&80.25&83.80&75.00&79.16\\
                DetectGPT&79.83&78.69&81.83&80.23\\
			GLTR&92.58&93.23&91.83&92.53\\
                BERT&94.55&93.59&95.74&94.65\\

                RoBERTa&95.87&94.35&\textbf{97.50}&95.90\\
			GPT2&93.75&93.25&94.33&93.79\\
                {\tool}\textsc{ without BigBird}&96.33&96.18&96.50&96.34\\
                {\tool}\textsc{ without Triplet Loss}&96.75&96.21&97.33&96.77\\		{\tool}&\textbf{97.67}&\textbf{98.64}&96.67&\textbf{97.64}\\
			\hline
		\end{tabular}
		\label{tab:baselines}
	\end{center}
	\vspace{-2mm}
\end{table}

\begin{table}
	\caption{Examples of Challenging Cases to Detect}
	\small
	\begin{center}

		\begin{tabular}{p{2.5cm}|p{5.5cm}}

		Post Question&Post Answer\\

			\hline
			How to directly set response body to a file stream in ASP.NET Core middleware?
			& ChatGPT-generated Answer: To set the response body to a file stream in ASP.NET Core middleware, you can use the Response.Body property. This allows you to write the file stream directly to the response, efficiently sending files to the client.\\
					\hline

				How to add shadow to ClipOval in flutter?
			& Human-written Answer: 	
	You can create your own CustomClipper:
\begin{verbatim}
class CustomClipperOval extends 
CustomClipper<Rect> \{
......
\end{verbatim}
\\
\hline

		\end{tabular}
		
		\label{tab:bad_example}
	\end{center}
	\vspace{-2mm}
\end{table}

While our \tool{} outperforms the baselines in most cases, there are instances where it may struggle to accurately identify ChatGPT-generated content.
One such scenario is when the ChatGPT-generated answer closely mimics human writing style, particularly when the answer is short.
As shown in the first example of Table~\ref{tab:bad_example}, the ChatGPT-generated answer is straightforward and directly addresses the question, closely mimicking a human response. 
Due to the lack of distinctive AI features in short answers, our \tool{} might misclassify these as human-generated. This human-like simplicity and directness can pose a challenge to accurate detection.

Another challenging case is when the human-written answer contains less natural language and includes long code snippets. 
These characteristics may mislead our model into classifying it as an AI-generated response. 
As shown in the second example of Table~\ref{tab:bad_example}, the human-written answer includes extensive code, which our \tool{} might incorrectly interpret as an AI-generated response due to the unnatural language patterns and structure. 
This misclassification highlights a limitation of our model in distinguishing between human and AI content when faced with answers heavily dominated by code and lacking in natural language cues.

\rqans{\textbf{Answer to RQ1:} The evaluation results clearly demonstrate the superior performance of our \tool{} tool compared to established baselines. 
Our model achieves the highest accuracy, precision, and F1-score among all tested models. While RoBERTa exhibits a slightly better recall score, the precision of our model surpasses all baselines, making it more effective in accurately detecting ChatGPT-generated content without misclassifying human-generated content. 
Additionally, our ablation study confirms the effectiveness of incorporating the BigBird model and triplet loss, as both modifications lead to significant improvements in performance metrics. 
This highlights the importance of these components in enhancing the capability of our model to handle long texts and leverage reference answers for more precise predictions.
}

\subsection{ RQ2: How does the length of the input text affect the performance of \tool{}?
}
\label{sec:rq2resuls}
\begin{figure*}
    \centering
    \includegraphics[scale=0.5]
    {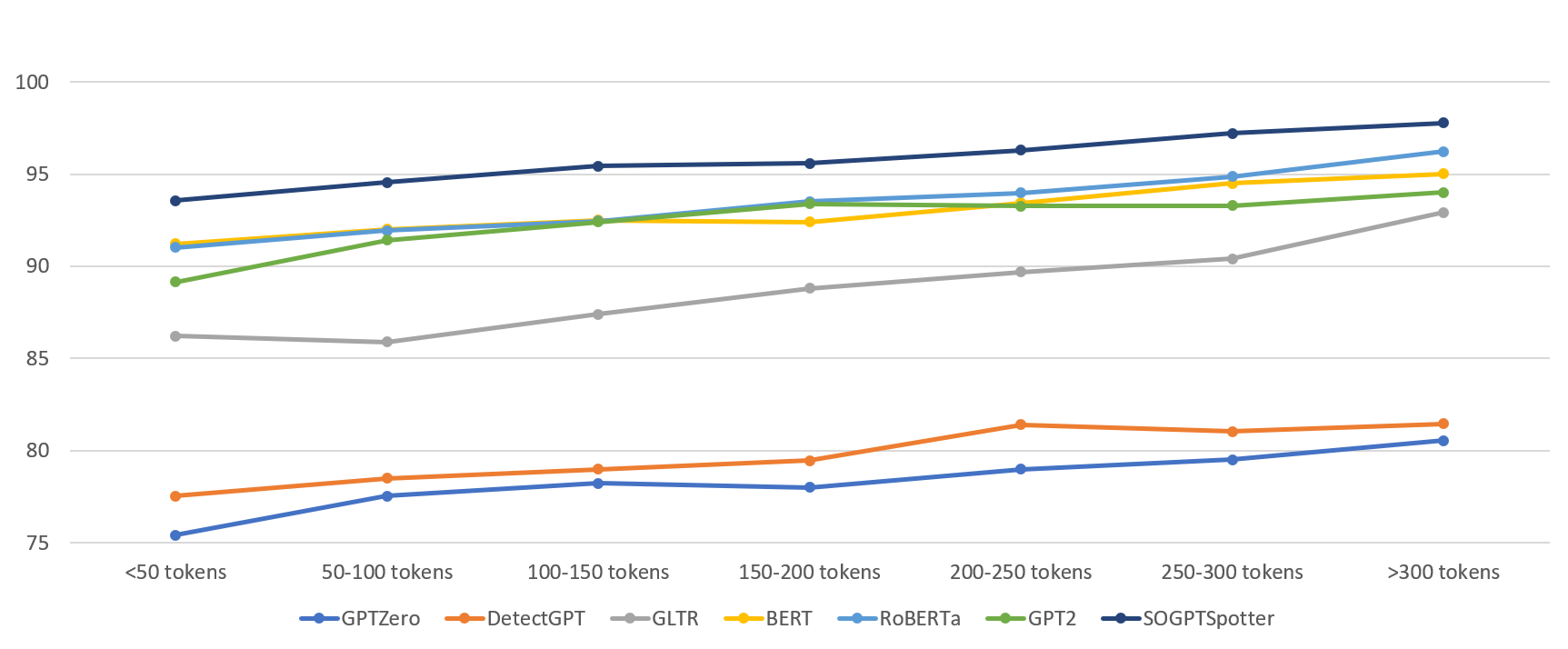}
    \caption{ Results for Different Approaches  with Different Token Lengths}
    \label{fig:length}
\end{figure*}

\begin{figure*}
    \centering
    \includegraphics[scale=0.5]
    {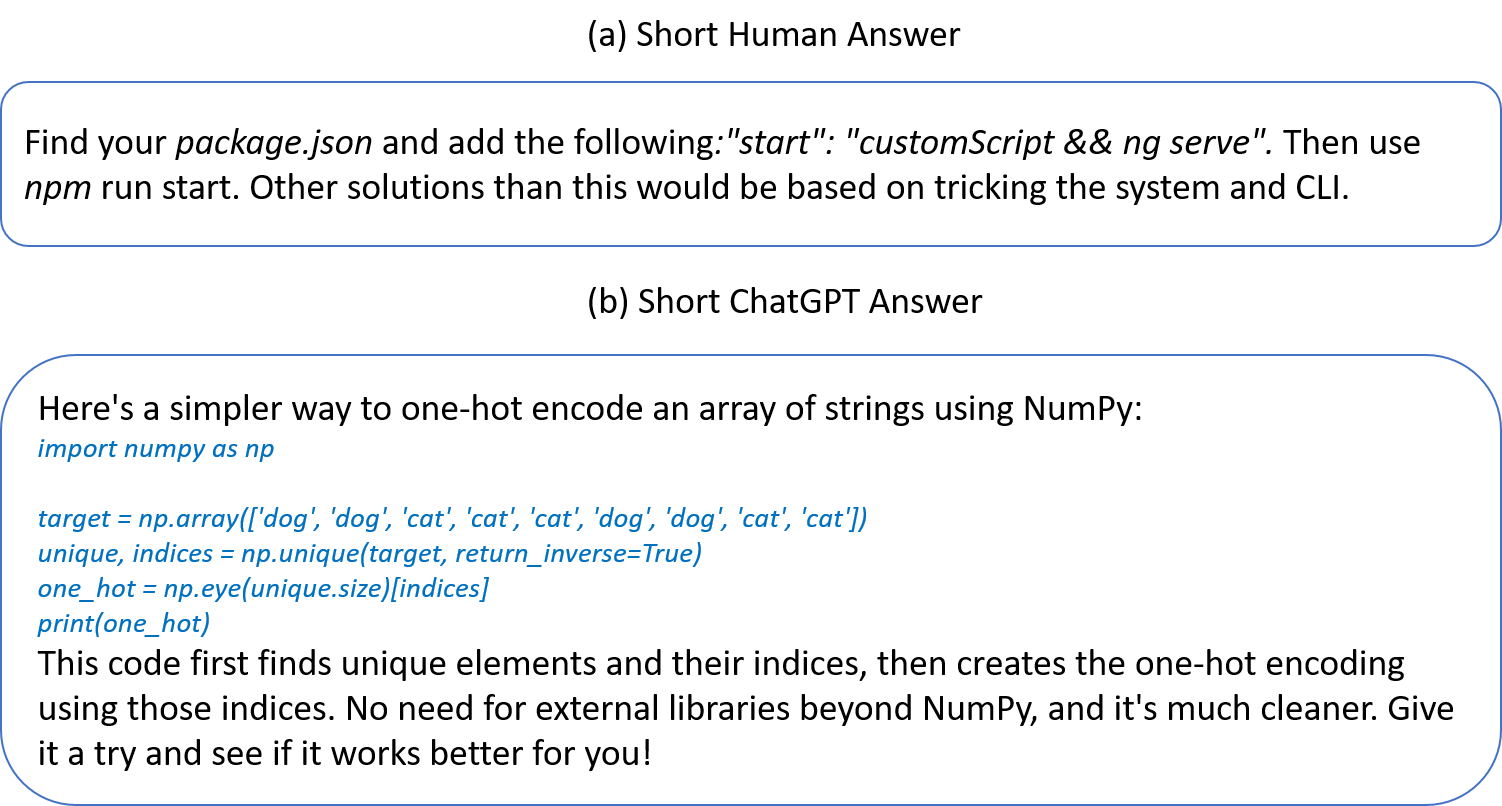}
    \caption{Examples of Short Post Answers}
    \label{fig:short}
\end{figure*}

The length of the input text significantly impacts the performance of machine learning-based approaches. Typically, longer texts provide more context and information, which helps in better detection accuracy.
We designed an experiment to investigate the impact of input text length on the performance of our tool and the baseline models. 
We varied the token length of the input text from less than 50 to more than 300 tokens and measured the F1-score of our tool and the baseline models at each token length.
By comparing the performance of the models at different text lengths, we were able to quantify the impact of input text length on the detection accuracy of each model.

Figure~\ref{fig:length} shows the performance of our tool and the baseline models across different token lengths.
As depicted, \textbf{all models exhibit improved F1-scores with longer input texts}.
Our tool demonstrates a consistent improvement in F1-score as the token length increases.
This suggests that longer texts, providing more context and information, enable our tool to more effectively distinguish between human and ChatGPT-generated responses.
The baseline models also demonstrate a similar trend, with their F1-score improving with longer input texts.

However, it is noteworthy that \textbf{our \tool{} consistently outperforms all the baseline models in all situations}, demonstrating its superior detection ability in a wide range of input lengths.
As shown in Figure~\ref{fig:short}, our model also excels at accurately classifying shorter answers that other baseline models fail to detect correctly.
In the figure~\ref{fig:short}, (a) is a short human answer within a 50 maximum token length, and (b) is a short answer from ChatGPT within a 100 maximum token length. 
We can see that they both lack adequate linguistic information due to the short length. 
All the other baselines fail to classify these correctly, but our model correctly classifies them because it can also utilize the information from the post question and the inner knowledge of the ChatGPT model.
The ability of \tool{} to handle both short and long texts efficiently makes it a reliable tool for diverse Stack Overflow posts. 


\rqans{\textbf{Answers to RQ2:} The length of the input text significantly influences the performance of \tool{}, with longer texts providing better F1-scores. 
Our tool consistently outperforms all baselines across different text lengths, highlighting its effectiveness and robustness in handling texts of varying lengths.
}

\subsection{RQ3: How robust is our \tool{} when subjected to adversarial attacks?}
\begin{table*}
	\caption{Results for Different Approaches with Adversarial Attacks}
	\small
	\begin{center}
		\begin{tabular}{l|c|c|c}
		    \hline
			Models &  synonym substitution& perturbation& paraphrasing\\
			\hline
			GPTZero&75.60&77.76&73.11\\
			DetectGPT&76.41&78.58&74.56\\
    		GLTR&87.66&89.55&87.25\\
      		BERT&89.55&91.88&91.16\\
                RoBERTa&89.94&92.03&92.33\\
   			GPT2&86.46&90.08&89.53\\
			{\tool}&94.43&94.90&95.85\\
			\hline
		\end{tabular}
		\label{tab:Res2}
	\end{center}
	\vspace{-2mm}
\end{table*}

A variety of adversarial attacks could impact the performance of \tool{} and baselines.
Such attacks include human editing ChatGPT-generated content to try and make it look more human-generated. 
To evaluate the robustness of our \tool{} under such attacks, we designed an experiment where we introduced several different adversarial attacks to the ChatGPT-generated answers.
These adversarial attacks included substitution~\cite{ren2019generating}, perturbation~\cite{gao2018black} and paraphrasing~\cite{iyyer2018adversarial}, which were applied to the answers to make them more challenging to detect.
The substitution method is a widely utilized adversarial attack method that efficiently performs synonym substitution based on word saliency scores and maximum word-swap variance.
The perturbation method employed is Deep-Word-Bug, which incorporates random substitution, swapping, deletion, and insertion, mimicking real-world human activities.
The paraphrasing method is an encoder-decoder model designed for syntactically controlled paraphrase generation.

For each type of adversarial attack, we ran \tool{} and the baseline models on our modified test dataset and measured their detection accuracy using F1-score. 
This allowed us to quantify the impact of each type of adversarial attack on the performance of our tool and the baseline models, providing valuable insights into their robustness against such attacks.
The results are shown in Table~\ref{tab:Res2}.

All three adversarial attack methods impacted the performance of our model and the baselines, making the detection task more challenging.
However, \textbf{our \tool{} consistently outperformed all the baseline models in each type of attack scenario.}
Specifically, our model achieved F1-scores of 94.43, 94.90, and 95.85 for synonym substitution, perturbation, and paraphrasing, respectively.
Compared to the baselines, our model showed a performance improvement of 24.91\% over GPTZero, 23.58\% over DetectGPT, 7.72\% over GLTR, 5.45\% over BERT, 4.99\% over RoBERTa, and 9.22\% over GPT2 for synonym substitution. 
For perturbation, our model showed a performance improvement of 22.04\% over GPTZero, 20.77\% over DetectGPT, 5.97\% over GLTR, 3.29\% over BERT, 3.12\% over RoBERTa, and 5.35\% over GPT2. 
For paraphrasing, our model showed a performance improvement of 31.10\% over GPTZero, 28.55\% over DetectGPT, 9.86\% over GLTR, 5.14\% over BERT, 3.81\% over RoBERTa, and 7.06\% over GPT2.

\textbf{In terms of performance degradation, our model exhibited the lowest decrease ratio compared to the baseline models.}
The degradation ratios for our model were 3.29\% for synonym substitution, 2.81\% for perturbation, and 1.83\% for paraphrasing, which are lower compared with all the baselines.
Specifically, in synonym substitution, GPTZero, DetectGPT, GLTR, BERT, RoBERTa, and GPT2 showed degradation of 4.50\%, 4.76\%, 5.26\%, 5.39\%,6.21\%, and 7.82\%, respectively.
In perturbation, the degradation for GPTZero, DetectGPT, GLTR, BERT, RoBERTa, and GPT2 were 1.77\%, 2.06\%, 3.22\%, 2.93\%, 4.04\%, and 3.96\%, respectively. For paraphrasing, the degradation observed for GPTZero, DetectGPT, GLTR, BERT, RoBERTa, and GPT2 were 7.64\%, 7.07\%, 5.71\%, 3.69\%, 3.72\%, and 4.54\%, respectively.

The results show that synonym substitution is particularly effective against the language models BERT, RoBERTa, and GPT2. 
This attack method reduced their F1-scores significantly.
The likely reason is that synonym substitution can significantly alter the surface form of the text, reducing the ability to distinguish humans from AI-generated features.

For GPTZero, DetectGPT, and GLTR, paraphrasing proved to be the most effective adversarial attack. The F1-scores for these models dropped significantly.
This is likely because these models rely heavily on statistical measures such as perplexity, which can be significantly affected by paraphrasing.

Our \tool{}, on the other hand, demonstrated robustness across all types of adversarial attacks. 
The F1-scores for our model were high for synonym substitution, perturbation, and paraphrasing.
The reason for this robustness is likely due to our model's use of the reference answer for comparison. 
This approach allows our model to retain a high level of accuracy even when the surface form of the text changes, as the core meaning of the content remains similar.

\rqans{\textbf{Answers to RQ3:} These results demonstrate the robustness of our \tool{} against adversarial attacks. Despite the introduction of attacks designed to make detection more difficult, our tool was able to maintain a high detection accuracy, outperforming the baseline models in all scenarios. 
However, its performance did degrade and other adversarial attacks may cause further degradation.
}

\subsection{RQ4: How well does \tool{} generalize across different domains and various LLMs?} 

\begin{table*}[htbp]
\caption{Generalization Results of \tool{} and Baselines Across Different Domains and LLMs}
\small
\begin{center}
\begin{tabular}{l|c|c|c|c|c|c}
\hline
 & \multicolumn{3}{c|}{Different Domains} & \multicolumn{3}{c}{Different LLMs} \\
 \hline
Models & Mathematics &  Electronics & Bitcoin & LLaMA 3 & Claude 3 & Gemini \\ 
\hline
GPTZero & 80.96 & 78.85 & 78.40 & 78.30 & 76.57 & 79.60 \\
DetectGPT & 79.23 & 82.55 & 79.62 & 80.34 & 78.30 & 83.90 \\
GLTR & 90.57 & 87.48 & 86.54 & 86.49 & 88.70 & 88.35 \\
BERT & 91.51 & 90.20 & 87.94 & 86.71 & 86.49 & 87.00 \\
RoBERTa & 92.00 & 90.57 & 89.40 & 88.70 & 87.00 & 90.57 \\
GPT-2 & 89.53 & 87.48& 88.40 & 86.49 & 88.35 & 85.85 \\
\tool{} & \textbf{94.49} & \textbf{93.48} & \textbf{92.47} & \textbf{92.97} & \textbf{91.24} & \textbf{92.47} \\
\hline
\end{tabular}
\label{tab:generalization}
\end{center}
\vspace{-2mm}
\end{table*}

To investigate the generalization of our \tool{}, we conducted two different studies.
Firstly, to examine the generalization of our tool across different domains, we collected 100 high-quality posts from three different websites: Mathematics Stack Exchange\footnote{https://math.stackexchange.com/},  Electronics Stack Exchange\footnote{ https://electronics.stackexchange.com/}, and Bitcoin Stack Exchange \footnote{ https://bitcoin.stackexchange.com/}. 
We used similar selection criteria and preprocessing methods as used in selecting our Stack Overflow posts.
The posts were selected from the Stack Exchange Data Explorer. 
For each domain, we used 100 post answers as human answers, generated ChatGPT answers and also generated reference answers using similar approach as we used in our dataset generation. 
This resulted in 100 (reference answer, human answer) pairs and 100 (reference answer, ChatGPT answer) pairs for each domain.

Secondly, to evaluate the performance of our tool with different LLMs, we selected 100 posts from the test dataset and generated non-human answers using Claude 3 Sonnet\footnote{ https://www.anthropic.com/api}, LLaMA 3 \footnote{https://llama.meta.com/llama3/}, and Gemini Flash \footnote{https://ai.google.dev/gemini-api}.
We used the similar techniques in our data generation  to ask these LLMs to generate the reference answers and AI answers.
This resulted in 100 (reference answer, human answer) pairs and 100 (reference answer, AI answer) pairs for each LLM. 
This provided us with a comprehensive evaluation dataset to assess the robustness and generalization capabilities of \tool{} across different domains and LLMs.

Table~\ref{tab:generalization} shows the results of our model and baselines cross different domains and LLMs.
In the cross domains experiment, GPTZero and DetectGPT showed similar results across different domains because they are pre-trained and not further fine-tuned. 
They rely more on the statistical features of the text rather than domain-specific nuances. 
Specifically, GPTZero's F1-score degradation ratios were -2.27\% in Mathematics, 0.39\% in Engineering, and 0.96\% in Bitcoin.
DetectGPT showed degradation ratios of 1.25\% in Mathematics, -2.89\% in Engineering, and 0.76\% in Bitcoin.
All the other models performed best in the Mathematics domain. 
In Mathematics domain, the F1-score degradation ratios is 2.12\%, 3.32\%, 4.07\%, 4.54\%, and 3.23\% for GLTR, BERT, RoBERTa, GPT2 and \tool{}.
The reason is that the posts in Mathematics often include LaTeX formulas, which are similar to the writing format in Stack Overflow. 
In the Electrical Engineering domain, the presence of electrical circuit images poses challenges for all models.
This is reflected in the degradation ratios, with GLTR showing a degradation ratio of 5.46\%, BERT at 4.70\%, RoBERTa at 5.56\%, and GPT2 at 6.73\%, \tool{} at 4.26\%. 
The Bitcoin domain's diverse financial content further impeded performance.
The F1-score degradation ratios in Bitcoin for GLTR, BERT, RoBERTa, GPT-2, \tool{} were  6.47\%, 7.09\%, 6.78\%, 5.75\%  and 5.29\%. 
\tool{} demonstrated a lower degradation ratio in both Electronics and Bitcoin domains, indicating it is slightly more robust compared to other models.
\textbf{What is more, the performance of our tool still highly outperform all these baselines in the cross domain settings.}
Our model's superior performance across these domains can be attributed to its unique structure leveraging reference knowledge. 
The reference answers generated by ChatGPT provide domain-specific insights, aiding in better detection accuracy.

For the different LLMs experiments, GPTZero and DetectGPT also showed similar results across different LLMs. 
Specifically, GPTZero's F1-score degradation ratios were 1.09\% for LLaMA 3, 3.27\% for Claude 3, and -0.56\% for Gemini.
DetectGPT showed degradation ratios of -0.14\% for LLaMA 3, 2.41\% for Claude 3, and -4.57\% for Gemini.
\textbf{The performance of all other models declined, but our model still outperformed the others and showed a lower degradation ratio.}
\tool{} achieved F1-scores of 92.97\% for LLaMA 3, 91.24\% for Claude 3, and 92.47\% for Gemini, compared to its original F1-score of 97.64\%.
This indicates a degradation of 4.78\%, 6.55\%, and 5.29\% respectively.
GLTR showed a  degradation of 6.53\%, 4.14\% and 4.52\%  for LLaMA 3, Claude 3 and  Gemini.
BERT showed a degradation of 8.39\%, 8.62\%, 8.08\% for LLaMA 3,  Claude 3, and  Gemini.
RoBERTa showed a degradation of 7.51\%, 9.28\%, 5.56\% for LLaMA 3,  Claude 3, and  Gemini.
GPT-2 showed a degradation of 7.78\%, 5.80\%, 8.47\% for LLaMA 3,  Claude 3, and  Gemini.
Our model's capability to utilize the knowledge of these LLMs by generating reference answers plays a crucial role in maintaining higher performance.
These results highlight the generalization capability of \tool{} across different domains and various LLMs, showcasing its effectiveness and robustness in diverse and challenging environments.

\rqans{\textbf{Answers to RQ4:} Our \tool{} demonstrates strong generalization capabilities across different domains and various LLMs. 
Despite the challenges posed by domain-specific content and different LLMs, \tool{} consistently outperforms baseline models, maintaining high detection accuracy with lower performance degradation.}

\subsection{RQ5: How effective is \tool{} in a real-world setting?}

\begin{figure}
    \centering
    \includegraphics[scale=0.5]{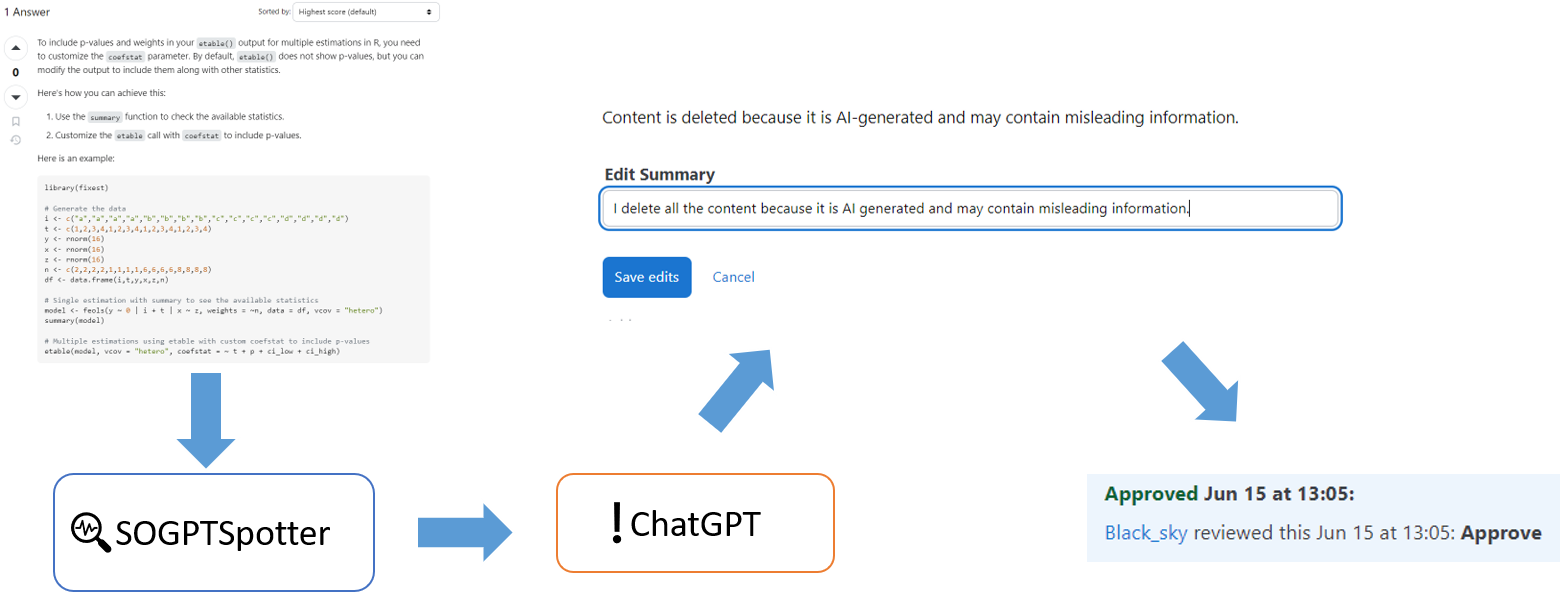}
    \caption{Example for using \tool{} in real case }
    \label{fig:real-case}
\end{figure}

Our \tool{} is aimed at aiding human moderators of Q\&A sites like Stack Overflow in detecting ChatGPT or similar AI-generated content.
We wanted to evaluate \tool{}'s potential for real-world post classification.
We conducted a small-scale field study to evaluate the practical application of our \tool{} in detecting ChatGPT-generated content on Stack Overflow. 
We randomly selected 50,000 post answers from the period between November 30, 2022, and April 30, 2024, ensuring that these posts were not included in our training dataset to avoid potential bias. 
Our model flagged 146 post answers as containing potentially AI-generated answers. 
From these, we selected 50 post answers for further manual review, which is a manageable size for human effort to manually submit the post edits.
In Stack Overflow, each question can have up to 5 tags to describe its topic. 
The 50 selected post answers contain 124 tags in total (we took tags from their corresponding question), and 69 of these tags are unique. 
This indicates that the 50 selected posts cover a diverse set of Stack Overflow topics.
Within the 50 selected post answers, 42 (84\%) of them are more than 50 tokens long, while the others are shorter answers.
And 38 (76\%) of these post answers contain code snippets, the others are pure text.

As shown in Figure~\ref{fig:real-case}, after our tool classifies a post answer as ChatGPT-generated, we further edit it by deleting all the content and leaving a comment like "This content is AI-generated and may contain misleading information" to notify the moderator. 
This process ensures that moderators are aware of the AI-generated nature of the content and can take appropriate action.
We submitted the 50 post edits to the community for approval. 
For the 50 submitted post edits, 47 (94\%) were accepted and these posts were further deleted.
Only 3 (6\%) were rejected. 
Among the three rejected posts, two were short answers.
These two posts were notably shorter than the others ( less than 30 tokens), which likely contributed to the challenge of accurately classifying them. 
For example, one post stated:
\begin{small}
"It happens sometimes because of Android Studio cache issues. Try to invalidate caches and restart the studio. In below image select all the options."
\end{small}
Such short text often lack the complexity and nuance that AI models struggle to replicate, making it harder to distinguish between human-written and AI-generated content (see RQ2 results in Section \ref{sec:rq2resuls}).
Another rejected post answer has long code snippets and short text. 
It stated:
\begin{small}
"In the code snippet above, \texttt{pil\_mask} is created as a PIL Image, then converted to a NumPy array (\texttt{pil\_mask\_np}). You can now use \texttt{pil\_mask\_np} wherever you need it in your code.
<long code snippet>."
\end{small}
Answers with extensive code snippets and minimal explanatory text can be misleading. 
These answers lack the nuanced language patterns that our tool uses to detect AI-generated content, leading to misclassification. 
However, for the 47 accepted post edits, the trusted contributors believed that they were AI-generated content and thus deleted them.
This real-world acceptance demonstrates the usefulness of our tool in spotting ChatGPT content on Q\&A collaborative sites.

\rqans{\textbf{Answer to RQ5:} The field study results demonstrate the practical effectiveness of our \tool{} tool in a real-world setting on Stack Overflow. 
Our model successfully identified ChatGPT-generated content with a high acceptance rate of 94\% for post edits submitted for community review.
Despite some challenges with shorter answers and posts containing long code snippets, the overall high acceptance rate indicates that our tool can significantly aid in moderating AI-generated content on Q\&A platforms.}

%% file: threats.tex
\section{Threats to Validity}

\textbf{ Internal Validity.} Threats to internal validity in our experiments may arise from parameter tuning. 
The choice of parameters for the model, such as the learning rate, the number of layers in the network, and the number of training epochs, can significantly affect the results. 
There is always a possibility that a different set of parameters could yield better results. 
We conducted preliminary experiments to find the optimal set of parameters for our model. 
Despite these efforts, there remains a potential bias introduced by the specific parameter settings chosen, and it is possible that alternative settings could further optimize the model's performance.

Another internal validity threat is the selection of post answers from Stack Overflow as human answers. 
There is no standard to evaluate the quality of non-AI features in these answers. 
Although Stack Overflow allows users to upvote answers, the ratings are primarily based on the overall content quality of the answer. 
This reliance on user votes might not accurately reflect the presence or absence of non-AI features, potentially introducing bias in our dataset.

\textbf{External Validity.} The main external threat to the validity of our work is the dataset generalization. 
Our study used a dataset of post-answer pairs from Stack Overflow, which is a popular Q\&A platform for technical questions. 
However, the characteristics of posts on Stack Overflow may not be representative of all Q\&A platforms. 
For instance, the language used, the length of posts, and the types of questions asked could vary across different platforms. 
Although the generalization experiments show the effectiveness of our model across different domains, future work could involve training the model on different Q\&A platforms to improve its generalizability and adaptability to various contexts and content types.


Another potential threat is the generalizability of our model to future versions of LLMs, as these models continue to evolve rapidly. 
While our \tool{} performed well on both the experimental datasets and real-world case studies, improvements in LLMs might lead to more sophisticated AI-generated content that is harder to detect. 
Our model's reliance on  AI characteristics of reference answer might limit its effectiveness against future advancements in AI content generation. 
Thus, it is crucial to continuously update reference answer generation to keep pace with the advancements in LLMs.

\textbf{Construct Validity.} A key aspect affecting construct validity is the prompt design for generating ChatGPT answers. 
Although we used various techniques to increase the diversity of ChatGPT-generated answers, such as different prompts, context length control, asking ChatGPT to mimic a human tone randomly, reducing indicative words, and randomly adding filler words, there are still scenarios we did not consider. 
For instance, in real use cases, people might generate a ChatGPT answer and then partially edit it with human-written content. 
This hybrid content presents a unique challenge that our current prompts and data generation techniques do not fully address. 
To improve the diversity and robustness of ChatGPT answers, future work should consider incorporating such mixed-content generation methods and other real cases. 

%% file: relatedWork.tex
\section{Related work}

\subsection{Stack Overflow Analysis}

Due to the popularity of Stack Overflow in the developers' community, numerous studies have looked at various aspects of Stack Overflow questions and answers. Several have investigated post quality, developed quality metrics, post recommendation and a few even looked at generating summaries or question titles. Various studies have looked at Stack Overflow post quality. Duijn et al looked into characterizing quality questions and code \cite{duijn2015quality}.  Mondal et al \cite{mondal2023automatic} investigated why edits are rejected for some Stack Overflow posts. Ponzanelli et al \cite{ponzanelli2014improving} developed mechanisms for classifying low quality Stack Overflow posts. Several studies have looked at characterizing the quality of questions and answers and recommending improvements. Opu and Roy \cite{opu2022towards} investigated techniques to improve the quality of Stack Overflow questions. Wang et al.\cite{wang2018users} investigated the post revision practices of Stack Overflow users, including quality enhancement. 

Various studies have used machine learning models for answer and code recommendation from Stack Overflow posts. Gao et al \cite{gao2023know} proposed an answer recommender using a deep learning-based ranking approach. The authors also proposed a question boosting-based approach for high quality answer recommendation \cite{gao2020technical}.  A small number of works have generated some aspects of Stack Overflow content.  Cai et al \cite{cai2019answerbot} proposed a bot that provides Stack Overflow post summarisation to aid developers in generating summarised content from existing Stack Overflow posts.   Gao et al \cite{gao2021code2que,gao2020generating} developed tools to generate high quality questions from code snippets to aid developers in producing posts with questions more likely to be answered. Only question tiles are generated and not post content. Similarly, Zhang et al \cite{zhang2022improving,zhang2023diverse} also propose approaches to generate high quality titles for posts using transformer models.

\subsection{AI-generated text detection}

The detection of AI-generated content has been a topic of increasing interest in recent years. 
This is due to the rise of sophisticated language models such as GPT-2, GPT-3, and ChatGPT, which can generate human-like text, leading to potential misuse in various domains, including online platforms like Stack Overflow.

To tackle the problem of identifying machine-generated text, researchers have employed a variety of methodologies. One of the initial and effective strategies is the use of a bag-of-words classifier, which has been empirically shown to differentiate between human-authored and GPT-2 generated texts~\cite{solaiman2019gpt2}. However, this approach encounters significant challenges when applied to more sophisticated versions of the GPT-2 model and complex sampling methodologies.

Zellers et al. \cite{zellers2019defending} proposed that the models best equipped for generating neural disinformation are, in turn, capable of recognizing their own output. However, it is important to note that a detector specifically engineered for a particular generator may lack the adaptability to function effectively with an alternate generator\cite{uchendu2020detection}.

Recognizing these limitations, the research community has shifted its focus towards RoBERTa~\cite{liu2019roberta}, a model that has demonstrated remarkable efficiency across diverse classification tasks relevant to the detection of machine-generated text. After the process of fine-tuning, RoBERTa has proven its prowess as a detector across multiple domains~\cite{solaiman2019gpt2, fagni2021detection, rodriguez2022crossdomain}.


Finally, it is worth mentioning that OpenAI has disseminated a detector trained on text derived from 34 distinct language models, using text obtained from Wikipedia, WebText~\cite{radford2019language}, and their proprietary human demonstration data. 






Another tool developed for detecting AI-generated text is GLTR~\cite{gehrmann2019gltr}.
It uses the distribution of the next possible words to detect if a text was written by a human or a machine. 
However, it has been shown that GLTR can be evaded by paraphrasing the AI-generated text~\cite{zellers2019defending}.
GPT-2 output detector is anothe detection tool developed by OpenAI. 
This tool is trained on a dataset of human-written text and GPT-2 generated text~\cite{Gpt2output}.
Recently, a new tool called GPTZero has been proposed~\cite{gptzero}.
The detector can make these predictions at various levels of granularity, including individual sentences, paragraphs, and entire documents.

\subsection{Authorship classification}
Authorship classification is another area of research that is relevant to our work.
Authorship classification is a well-studied field in computational linguistics and has been applied to various domains, from identifying the authors of classical texts to detecting fraudulent content online.
The task involves attributing a given piece of text to one of several potential authors based on stylistic and linguistic features.

One of the earliest works in this field was conducted by Mosteller and Wallace~\cite{mosteller1964inference}, who used statistics to attribute the disputed Federalist Papers to their likely authors. 
They used function words, which are words that have little lexical meaning but serve to express grammatical relationships, as features for their model. 

In recent years, machine learning techniques have been widely used for authorship classification.
For instance, Stamatatos~\cite{stamatatos2009survey} provides a comprehensive survey of various machine learning methods used for authorship attribution, including Naive Bayes, Support Vector Machines (SVM), and neural networks.
These methods typically involve training a model on a set of known-authorship texts, and then using this model to predict the authorship of unknown texts.
A significant work in the field was done by Koppel et al.~\cite{koppel2009computational}, who proposed a method based on 'impostors'. In this approach, a set of candidate authors is chosen, and each is represented by a set of texts. 
To attribute a given text, it is compared to each candidate's texts, and the candidate whose texts are most similar to the given text is chosen as the author.
Another notable work is by Seroussi et al.~\cite{seroussi2012authorship}, who used LDA for authorship attribution and achieved high accuracy on several datasets.

However, authorship classification methods have their limitations. 
They typically require a large amount of known-authorship text for training, and their performance can be affected by various factors such as the length of the texts and the number of candidate authors. ~\cite{hossain2021authorship}.




%% file: conclusion.tex
\section{Conclusion}
In this study, we introduced {\tool}, a novel approach for detecting ChatGPT-generated answers on Stack Overflow. 
Our approach leverages BigBird-based Siamese Neural Networks with triplet loss and was trained using a dataset of triplets, comprising reference answers, human answers, and ChatGPT-generated responses.
Our empirical evaluation revealed that \tool{} outperforms well-established baselines, including GPTZero, DetectGPT, GLTR, BERT, RoBERTa, and GPT-2, in identifying ChatGPT-synthesized responses.
We also conducted an ablation study to demonstrate the effectiveness of our model's components.
We conducted additional experiments to assess various factors such as the impact of text length on detection performance, robustness against adversarial attacks, and generalization across different domains and LLMs.
Furthermore, we performed a real-world case study on Stack Overflow to analyze the practical applicability of our tool.

Our findings suggest that factors like text length significantly influence the detection accuracy of such tools. 
Despite adversarial attacks designed to make detection more difficult, our \tool{} demonstrated robust performance, maintaining high detection accuracy across all scenarios.
The results from generalization and real-world case studies further underscore the effectiveness and practicality of our proposed approach.

In future work, we aim to enhance the robustness and adaptability of \tool{}.
This includes refining the encoder model to improve detection accuracy, improving its applicability to other language models and Q\&A platforms by designing more effective reference answer prompts and investigating its potential for real-time detection of AI-generated content. 
We also plan to study the social and collaborative implications of deploying our tool on various online Q\&A platforms.